\documentclass[aps,prd,superscriptaddress,preprint]{revtex4}
\usepackage{float}
\usepackage{amsmath,amssymb}
\usepackage[dvips]{graphicx}
\DeclareGraphicsExtensions{.jpg,.mps,.pdf,.png,.eps} 
\usepackage{color}

\newcommand{\dip}{Dipartimento di Fisica, Universit\`a di Trento\\
                                     via Sommarive 14, 38123 Trento, Italia\\}
\newcommand{\infn}{TIFPA (INFN)\\via Sommarive 14, 38123 Trento, Italia\ \medskip}

\newcommand{\guido}[1]{Guido Cognola${}^{#1}$\footnote{e-mail:\sl cognola@science.unitn.it\rm}}
\newcommand{\maxR}[1]{Massimiliano Rinaldi${}^{#1}$\footnote{e-mail:\sl massimiliano.rinaldi@unitn.it\rm}}
\newcommand{\luciano}[1]{Luciano Vanzo${}^{#1}$\footnote{e-mail:\sl vanzo@science.unitn.it\rm}}

\newcommand{\beq}{\begin{equation}}
\newcommand{\eeq}{\end{equation}}

\newcommand{\be}{\begin{equation}}
\newcommand{\ee}{\end{equation}}

\newcommand{\bea}{\begin{eqnarray}}
\newcommand{\eea}{\end{eqnarray}}

\newcommand{\ca}[1]{{\cal #1}}         
\def\nn{\nonumber}            
\def\eeq{\end{eqnarray}}      
\def\beq{\begin{eqnarray}}
               
\def\at{\left(}               
\def\aq{\left[}

\def\ct{\right)}              
\def\cq{\right]}

\def\R{{\hbox{{\rm I}\kern-.2em\hbox{\rm R}}}}   
\def\H{{\hbox{{\rm I}\kern-.2em\hbox{\rm H}}}}   
\def\N{{\hbox{{\rm I}\kern-.2em\hbox{\rm N}}}}   
\def\C{{\ \hbox{{\rm I}\kern-.6em\hbox{\bf C}}}} 
\def\Z{{\hbox{{\rm Z}\kern-.4em\hbox{\rm Z}}}}   
\def\ii{\infty}                                  
    
\def\al{\alpha}\def\be{\beta}

\def\la{\lambda}
\def\si{\sigma}\def\om{\omega}
\def\th{\theta}

\def\De{\Delta}\def\La{\Lambda}
\def\Om{\Omega}
\def\Th{\Theta}

\begin{document}

\title{Scale-invariant rotating black holes in quadratic gravity}

\author{\guido{}} \author{\maxR{}} \author{\luciano{}} 
\affiliation{\dip}
\affiliation{ \infn}

\date{\today}

\begin{abstract} 
\noindent Black hole solutions in pure quadratic theories of gravity are interesting since they allow to formulate a set of scale-invariant thermodynamics laws. Recently, we have proven that static scale-invariant black holes have a well-defined entropy, which characterizes equivalent classes of solutions. In this paper, we generalize these results and explore the thermodynamics of rotating black holes in pure quadratic gravity.
\end{abstract}

 \maketitle

\section{Introduction}

\section{Introduction}

\noindent The laws of black hole thermodynamics were established more than four decades ago by the seminal works of Carter, Bekenstein, and Hawking 
\cite{bekhawk,Bekenstein,Hawking-1974,Hawking-cmp1975,Hawking-cmp1976}. Despite many efforts, however, the connection between the macroscopic thermodynamical quantities, in particular the entropy, and the microscopic state counting is still an unsolved puzzle, although interesting proposals have been made in string theory \cite{StTh}, holography \cite{Holo}, and loop quantum gravity \cite{LQG}.

The common idea behind all these theories is that the fundamental origin of entropy and black hole evaporation must be looked for in the realm of quantum gravity, or at least of some semi-classical limit thereof, as proposed for the first time by Hawking. Since a consistent quantized version of general relativity is not available yet, the microscopic origin of the thermodynamical properties of black hole remains obscure.

In connection with these facts, most researchers believe that the road to the quantization of gravity inevitably goes through an extension of general relativity at high energy or at large curvature even at the \emph{ classical level}. Such extensions must require diffeomorphism invariance and equations of motion of at most the second order to be acceptable and there are several examples of theories with these characteristics. Here, we focus on a family of modified gravity models characterized by the replacement of the linear Einstein term in the action by a generic function $f(R)$ (for a review see e.g. \cite{defelice}). The most famous example  is the Starobinsky inflationary model, where the gravitational Lagrangian has the form ${\cal L}= R+R^{2}/(6\mu^{2})$, where $\mu$ is a  mass scale. The term quadratic in $R$ is added to describe, phenomenologically, the effects of quantum corrections on the metric $g_{\mu\nu}$ at large curvatures \cite{staro}. This strikingly simple extension of gravity is able to reproduce the data from cosmological observations with great accuracy and this might be a hint that research in $f(R)$ theories of gravity is a promising avenue.

Among the $f(R)$ models of gravity, the quadratic ones play a special role as they are believed to be renormalizable and asymptotically free  \cite{stelle,buchb,adler,tomb1,tomb2,frad1,frad2,avra,buch}. In general, these models all contain ghosts, except for  the simplest case of $f(R)=R^{2}$ \cite{kou}. The latter can be seen as the high curvature limit of the Starobinsky model and has attracted some interest in the context of supergravity \cite{sugra,ferrara} and early cosmology, where   it has been shown that deformations of $R^{2}$ induced by loop quantum corrections can support an inflationary  phase \cite{inflation,rcvz2,rcvz3}.

The pure quadratic model is the simplest example of scale-invariant theory. By adding other scale invariant operators in the action, as in \cite{strumia,kann}, one can generalize the theory to include also the standard model fields, which are particularly relevant in inflation.  In the present paper, we focus instead on the black hole solutions in $R^{2}$ gravity and the associated thermodynamics. In \cite{tn1} (see also \cite{riotto}) we have considered static topological black holes in $R^{2}$ with de Sitter and anti-de Sitter asymptotic geometry and found scale-invariant thermodynamical laws. In particular, by using the Euclidean approach, we have shown the Wald's formula still applies, provided one adjusts the sign in order to have a positive entropy. Here we study the asymptotically anti-de Sitter rotating solution of the same theory and generalize the results of our previous work. 

The Kerr-Anti de Sitter (KAdS) in Einstein gravity, first discovered by Carter \cite{carter:1968}, has been studied extensively in several aspects by many authors \footnote{An early attempt to describe the extremal branch of the solution can be found in \cite{Agnese:1999df}.}, notably in relation to supersymmetry \cite{Caldarelli:1998hg}, thermodynamics and CfT \cite{Dehghani:2001af,Caldarelli:1999xj,Cai:2005kw,Altamirano:2014tva}, AdS/CfT duality \cite{Skenderis:2000in,hawking:1998}, quasi-normal modes \cite{Gannot:2012pb,Cardoso:2013pza}, higher dimensions \cite{Awad:2000aj,Deser:2005jf,Madden:2004ym}, the theory of covariant conservation laws \cite{Barnich:2004uw,barn} and so on. In the context of $R^2$-gravity, and to the best of our knowledge, no results are available on KAdS black holes and we intend to fill this gap in the present work.

In the next section, we briefly describe the main features of $R^{2}$-gravity. In section 3 we review the Kerr-Anti de Sitter (KAdS) metric. Then we study it in the context of quadratic gravity in section 4, where we also present the laws of thermodynamics. Finally, we conclude with some considerations in section 5.

\section{Equations of motions}

\noindent Let us consider the Lagrangian density
\beq\label{eq1:L}
{\cal L}=\frac{\chi}{16\pi} \sqrt{\det{|g|}}R^{2}\,,
\eeq
where $\chi$ is a dimensionless parameter, which will play no role in what follows since we only consider the vacuum theory, and we arbitrarily set $\chi=1$. This expression is  part of the more general three-parameter quadratic gravity Lagrangian considered in \cite{stelle}, with the special property of being scale invariant, i.e. invariant under the substitution rule $g_{ik}(x)\to\la^2g_{ik}(x)$. The action is also invariant under $g^{'}_{ik}(x)=g_{ik}(\la^{-1}x)$ (that is not the diffeomorphism gauge symmetry), which means that we can omit the action of the scaling group on the coordinates in the substitution rule, showing that the action is equivalent to a pure Weyl scaling. Due to this symmetry, the dynamics based on \eqref{eq1:L} can only determine the light cone at each space-time point, so the conformal factor in the metric is not determined by the field equations. 

The vacuum field equations read
\beq\label{eq2:E}
R(R_{ik}-\frac{1}{4}Rg_{ik})-\nabla_{i}\nabla_{k}R+g_{ik}\nabla^2 R=0\,,
\eeq
which implies that $R$ obeys the massless wave equation
\beq
\nabla^2 R=0\,.
\eeq
We note that all Einstein spaces (i.e. metrics such that $R_{ik}=\La g_{ik}$ for some constant $\Lambda$), are solutions to the field equations. In contrast to standard general relativity (where Einstein spaces are obtained as solutions of the Lagrangian $ \sqrt{\det{|g|}}(R-2\Lambda)$ with a fixed  $\Lambda$), here the parameter $\La$  is a mere integration constant with no fundamental meaning. This tells us that  the spectrum of the theory contains all black hole solutions discovered over the years with flat, de Sitter or anti-de Sitter asymptotic geometry and flat, spherical or hyperbolic topological horizons. Does the appearance of the cosmological constant  mean that we have broken the scale invariance symmetry by some boundary condition? If we fix $\La$ this seems to be the case indeed, but if we allow $\La$ to vary then scale invariance is fully recovered, as shown in \cite{tn1}. This is similar to the breaking of rotational invariance by the gravitational field of the Earth: as long as it is considered an external field, it breaks the rotational symmetry, but once it is included into  the theory as a dynamical field,  the symmetry is restored. In particular, if we allow $\La$ to vary as a thermodynamical state space variable, the black hole spectrum becomes partitioned into orbits of the scale symmetry, which is then effectively mapped into the phase space of the theory. Generally speaking, scale symmetry is not spontaneously broken by the background (e.g. the  vacuum) in the classical theory. However, it is certainly broken in the quantum theory.

In the following, we shall need the Euclidean continuation of the action $I=\int{\cal L}$, which is given by the same expression as Eq.~\eqref{eq1:L}, except that $g_{jk}$ has no negative eigenvalues and $R$ is to be computed accordingly. The Euclidean equations of motion are identical to \eqref{eq2:E}, except that the d'Alembertian operator $\nabla^2$ is  replaced by the four-dimensional Laplace operator.

\section{The Kerr-AdS metric and its Euclidean continuation}

\noindent Let us know briefly review the metric describing a rotating black hole in the presence of a cosmological constant. In Boyer-Lindquist coordinates, first discovered by Carter \cite{carter:1968}, the metric reads \footnote{Here, we use the mostly plus signature.}
\beq\label{eq3:K}
ds^2 & = &-\frac{\De}{\rho^2}\at dt-\frac{a\sin^2\th}{\Xi}d\phi\ct^2+\frac{\De_{\th}}{\rho^2}\sin^2\th\at adt-\frac{a^2+r^2}{\Xi}d\phi\ct^2+\frac{\rho^2}{\De}dr^2 \\
&+& \frac{\rho^2}{\De_{\th}}d\th^2\nn\,,
\eeq
where
\beq\label{eq4:P}
\De=(r^2+a^2)(1+r^2/\ell^2)-2Mr, \quad \rho^2=r^2+a^2\cos^2\th\,,
\eeq
\beq
\De_{\th}=1-\frac{a^2}{\ell^2}\cos^2\th,\quad \Xi=1-\frac{a^2}{\ell^2},\quad \ell^2=-3/\La\,.
\eeq
We may think of $\ell$ as a cosmological scale. It is easy to see that, for $M>0$, any real root of $\De(r)=0$ must be positive and that there are at most two real roots. We impose the restrictions $|a|<\ell$, $\De^{'}(r_{+})>0$, where $r_{+}$ is the largest positive root of $\De(r)=0$. \\
The metric has the unique Killing field which is regular and null (namely $\zeta\cdot\zeta=0$) on the horizon
\beq\label{Kill}
\zeta=\partial_{t}+\Om_h\partial_{\phi}\,,
\eeq
where 
\beq\label{eq7:o}
\Om_h=\frac{\Xi a}{r_{+}^2+a^2}\,,
\eeq
is the angular velocity of the horizon. The metric is locally asymptotic to AdS$_{4}$ in a rotating frame with angular velocity $\Omega_{\infty}=-a/\ell^2$. This means that the solution rotates at infinity  with this angular velocity with respect to AdS$_{4}$ in the standard static coordinates. 
Thus the angular velocity of the black hole with respect to anti-de Sitter space is $\Om=\Om_h-\Om_{\ii}$.
Accordingly, we now show that the background metric with $M=0$ coincides with that of  AdS$_{4}$ in rotating coordinates. Let us take the AdS$_{4}$ metric in the form
\beq
ds_{b}^2=-\at1+\frac{Y^2}{\ell^2}\ct dT^2+\at1+\frac{Y^2}{\ell^2}\ct^{-1}dY^2+Y^2(d\Theta^2+\sin^2\Theta d\Phi^2)\,.
\eeq
The coordinate transformations \cite{hawking:1998}
\beq
T=t,\quad \Phi=\phi-at/\ell^2,\quad Y\cos\Theta=r\cos\th,\quad Y^2=\frac{r^2\De_{\th}\,,+a^2\sin^2\th}{\Xi}
\eeq 
map the AdS metric to Eq.~\eqref{eq3:K} with $M=0$. Note that points with constant angular coordinate in the $\phi$-frame rotate with respect to the $(T,Y,\Th,\Phi)$ global frame with angular velocity $\Om_{\infty}=-a/\ell^2$, as anticipated above.  
We now rotate the metric into Euclidean coordinates by performing the continuation $t=-i\tau$, $a=i\al$, which is needed  to make the differential $adt$ real. As a result, the metric takes the Euclidean signature in the region  $r>r_{+}$, where $\De>0$ and where it reads
\beq\label{eq5:K}
ds_{E}^2 & = &\frac{\De}{\rho^2}\at d\tau+\frac{\al\sin^2\th}{\Xi}d\phi\ct^2+\frac{\De_{\th}}{\rho^2}\sin^2\th\at \al d\tau-\frac{r^2-\al^2}{\Xi}d\phi\ct^2+\frac{\rho^2}{\De}dr^2 \\
&+& \frac{\rho^2}{\De_{\th}}d\th^2\nn\,.
\eeq
The substitution $a=i\al$ is understood also in the coefficients $\Delta$, $\Delta_{\theta}$, and $\Xi$. Note that $r_{+}>\al=|a|$ but that there are also some topological restrictions that must be applied in order to get a regular metric. A closer scrutiny shows that the sphere $r=r_{+}$ is a bolt in the manifold (i.e. a fixed point set of the co-rotating Killing field $\zeta=\partial_{\tau}-i\Om\partial_{\phi}$), around which the metric is periodic in $\tau$ with period 
\beq\label{eq6:b}
\be_{\tau}=\frac{4\pi\,(r_{+}^2+a^2)}{\De^{'}(r_{+})}=\frac{4\pi\ell^2\,(r_{+}^2+a^2)}{3r_{+}^3+(a^2+\ell^2)r_{+}-a^2\ell^2/r_{+}}\,.
\eeq
It can be shown that the quantity $\kappa=2\pi/\be_{\tau}$ is the surface gravity of the horizon (in the Lorentzian section) for a stationary observer, defined as an observer co-rotating with the horizon, in the limit $r\to r_{+}$.  
Therefore in the Euclidean section we identify $\tau\sim\tau+\beta_{\tau}$. One can also say that the Kerr-AdS metric is periodic in imaginary time with this period $\be_{\tau}$.  A further regularity condition is necessary. To see this, note that the metric on the sub-manifold  ${\cal H}=S^2(\rm{bolt})\times S^1$ (the Euclidean axis of symmetry) is
\[
d\si^2=\frac{\De_{\th}}{\rho_{+}^2}\al^2\sin^2\th d\om^2+\frac{\rho_{+}^2}{\De_{\th}}d\th^2\,,
\]
where $d\om$ is the local differential form  
\[
d\om= d\tau-\frac{r_{+}^2-\al^2}{\Xi\al}d\phi=d\tau+i\Om_h^{-1}d\phi\,.
\] 
Since the period of $\phi$ is $2\pi$ and that of $\tau$ is $\be_{\tau}$, by integrating over a cycle gives the condition
\[
\be_{\om}=\al\be_{\tau}-\frac{r_{+}^2-\al^2}{\Xi}2\pi=\al\at\be_{\tau}+\frac{2\pi i}{\Omega_h}\ct\,, 
\]
where $\be_{\om}$ is the period of $\om$. Hence there is a further identification, $\om\sim\om+\be_{\om}$. The quantity $\phi^{'}=\phi-i\Omega_h\tau$ is constant on the orbits of the Killing field, that is $\zeta\phi^{'}=0$, and like $\phi$ has period $2\pi$. The area of the two-dimensional bolt in a section of constant $\tau$ is
\beq\label{eq8:a}
A_{h}=\frac{4\pi(r_{+}^2-\al^2)}{\Xi}=\frac{4\pi(r_{+}^2+a^2)}{\Xi}\,.
\eeq
For $a$ real, the last expression is the area of the horizon within the Lorentzian manifold.

\section{Thermodynamics}

\noindent It is commonly accepted that in AdS space the canonical partition function is a well defined quantity.  By standard arguments, if we were to employ ordinary units in the loop expansion, the dominant tree level contribution, of order $\hbar^{-1}$, is given by the value of the classical Euclidean action, according to the semi-classical formula
\beq\label{eq9:Z}
I_{E}=-\log Z=\frac{1}{16\pi}\int_{K_{4}} R^2\sqrt{g}d^4x-\frac{1}{16\pi}\int_{b} R_{b}^2\sqrt{g_{b}}d^4x+\rm{boundary\;terms}\,,
\eeq
where the suffix $b$ indicates the reference background and $K_{4}$  the Euclidean Kerr-AdS metric. The boundary terms specify the boundary conditions, which define the statistical ensemble to be employed. This is also related to the question on which variables the partition function actually  depends, since this too is a matter of boundary conditions in the functional integral. The boundary of K$_4$ is a manifold $B=S^2\times S_{\be}^1$, where $\be$ (formerly called $\beta_{\tau}$) is the period of $\tau$  and the radius of $S^2$ is a certain value $r$ of the radial coordinate. We shall employ the canonical boundary conditions in which the boundary action is
\[
I_{B}=\frac{1}{8\pi}\int_{B}R(K-K_{b})\sqrt{h}d^3x\,,
\]
where $h_{ab}$ is the boundary metric, which is assumed to be the same as that of the background. By an adaptation of the Brown-York functional integral formalism for gravitational thermodynamics \cite{Brown:1992bq,by2,by3}, one sees that the ensuing action is appropriate for fixing the period $\beta$ and the angular velocity $\Om_h$, together with the two-metric on the boundary sphere and, of course, $\ell$ (which is kept fixed in the variational principle, anyway). Therefore we are dealing with the canonical ensemble and we should write the partition function as a function $Z(\be,\Om_h,\ell)$.  As we will see, since the background solution is a rotating AdS metric with angular velocity $\Om_{\ii}=-a/\ell^2$, the most relevant quantity to consider is the angular velocity relative to infinity
\beq
\Om=\Om_h-\Om_{\ii}=\Om_h+\frac{a}{\ell^2}\,,
\eeq
which we shall use in the following. Now for the Kerr-AdS solution $R=4\La=-12/\ell^2$ and the boundary action vanishes in the infinite volume limit,  so the bulk action reduces to the difference between the volume of the region $r>r_{+}$ and the one of the background enclosed by the spherical surface with radius $r$. By using the relation
\[
\sqrt{g}=\frac{\rho^2\sin^2\th}{\Xi}\,,
\]
a straightforward calculation gives
\beq\label{eq10:P}
\log Z=-\frac{\La^2}{3\Xi}\be\aq r_{+}^3+a^2r_{+}-M\ell^2\cq=-24\pi\,\frac{(r_{+}^2+a^2)^2(r_{+}^2-\ell^2)}{\ell^2(\ell^2-a^2)\,r_{+}\De^{'}(r_{+})}\,.
\eeq
To do this we rescaled the imaginary time of the background the amount $\tau\to(1-M\ell^2/r^3)\tau$ to match the metric of the background on the boundary surface to that of the actual solution,  which gives the last factor in Eq.~\eqref{eq10:P}. This result should be compared with the partition function of $K_4$ in Einstein gravity, which reads \cite{hawking:1998} ($G$ is the Newton constant)
\beq\label{eq11:P}
\log\tilde Z=-\frac{\pi(r_{+}^2+a^2)^2(r_{+}^2-\ell^2)}{G(\ell^2-a^2)\,r_{+}\De^{'}(r_{+})}
\eeq
Apart from $G$ this has the same dependence on the black hole parameters as for $Z$, but unlike $Z$ it is not scale invariant. Note that $M$ really is a length scale since we have no analogue of the Planck mass here. The free energy
\beq
F(\be,\Om,\ell)=-\beta^{-1}\log Z=\frac{24\pi}{\beta}\frac{(r_{+}^2+a^2)^2(r_{+}^2-\ell^2)}{\ell^2(\ell^2-a^2)\,r_{+}\De^{'}(r_{+})}
\eeq
is negative for black hole with sub-cosmological scale $r_{+}<\ell$, and positive otherwise, that is smaller black holes have less free energy than AdS space at the same temperature and therefore are thermodynamically favored. But for sufficiently large $r_{+}$ it is greater than AdS, indicating a phase transition is taking place on passing from low mass to higher mass black holes. 
 
We are ready to apply the thermodynamical rules to deduce the state functions of the black hole. Before this it is useful to deduce some general relation among them which stems only from the underlying scale invariance. Looking at \eqref{eq10:P} it is immediate to deduce the scaling rule $I_{E}(\la\be,\la^{-1}\Om,\la\ell)=I_{E}(\be,\Om,\ell)$. It then follows that
\beq
\frac{dI_{E}}{d\la}_{|\la=1}=0\,,
\eeq
that is
\beq
\be\frac{\partial I_{E}}{\partial\be}-\Om\frac{\partial I_{E}}{\partial\Om}+\ell\frac{\partial I_{E}}{\partial\ell}=0\,.
\eeq
If we introduce a ``cosmological volume'' $V\propto\ell^3$ and a pressure $P=-\partial I_{E}/\partial V$, we find the relation $\ell\partial I_{E}/\partial\ell=3V\partial I_{E}/\partial V=-3PV$, and we obtain a formula which can be interpreted as the equations of state of the black hole and its surrounding stuff,
\beq
3PV=\frac{m+\Om J}{T}\,,
\eeq
where $m$ is the thermodynamical internal energy and 
\beq
J=-\be^{-1}\frac{\partial I_{E}}{\partial\Om}= -\frac{\partial F}{\partial\Om}\,,
\eeq
is defined as the total angular momentum. 

We can do similar considerations for the entropy. It is convenient to regard the entropy as a function of the thermodynamical mass $m$, the angular velocity, and the cosmological scale $\ell$ (instead of the boundary data that define the functional integral of the partition function).  This is natural since the entropy is the value of the microcanonical partition function, which is computed by fixing the total energy in the functional integral, instead of the boundary metric \cite{Brown:1992bq}. 

The entropy in a scale-invariant theory is a scale-invariant concept (in contrast to conventional theories where it scales with volume, or horizon area in the case of black holes), that is
\beq	
S(\lambda^{-1}m,\lambda ^{-1}\Omega,\lambda\ell)=S(m,\Omega,\ell)\,,
\eeq
where $\la$ is the scaling parameter. It follows, as before, that
\beq
-m\frac{\partial S}{\partial m}-\Omega\frac{\partial S}{\partial\Om}+\ell\frac{\partial S}{\partial\ell}=0\,.
\eeq
We understand that
\[
m\frac{\partial S}{\partial m}=\frac{m}{T},\quad \frac{\partial S}{\partial\Om}=\frac{J}{T},
\]
where $J$ is the angular momentum, but we will not need this additional information in the following. Then, we have
\[
T\ell\frac{\partial S}{\partial\ell}=m+T\Omega\frac{\partial S}{\partial\Om}\,,
\]
and, finally,
\beq
TdS=dm+T\frac{\partial S}{\partial\Om}d\Om+\left(m+T\Om\frac{\partial S}{\partial\Om}\right)\frac{d\ell}{\ell}\,.
\eeq
This relation must be generally true as a consequence of the scaling rules and we note that it can be written in the manifestly scale invariant form
\beq
(\ell T)dS=d(\ell m)+T\frac{\partial S}{\partial\Om}d(\ell\Om)\,.
\eeq
We now must find the explicit expression for the mass, the entropy, and the angular momentum. With the Euclidean action method, together with  the results in the previous section, we first find the internal energy

\begin{eqnarray}
m&=& \frac{\partial\,I_{E}}{\partial\beta}=\frac{16\pi{r_{+}}\left(\beta\left(\ell^2-{r_{+}}^2\right)+4\pi \ell^2{r_{+}}\right)}{\beta\left(\ell^4-3 \ell^2{r_{+}}^2\right)+4\pi \ell^4{r_{+}}}
=\frac{16\pi{r_{+}}\left(\ell^2+{r_{+}}^2\right)}{\ell^4-a^2 \ell^2}\,,
\end{eqnarray}
which is related to the metric parameter $M$ by
\[
m=\frac{24\pi r_{+}^2}{\ell^2\Xi(r_{+}^2+a^2)}\,M\,.
\]
From this, we compute the entropy
\begin{eqnarray} 
S&=& \beta\,m-I_{E} = \frac{24\pi^2\left(a^2+{r_{+}}^2\right)}{\ell^2-a^2} = \frac{1}{4}|f'(R)|A_h\,,
\end{eqnarray}
and the angular momentum
\begin{eqnarray}
J&=& \be^{-1}\frac{\partial\,I_{E}}{\partial\Omega}=M_{e}\,a\,,
\end{eqnarray}
where 
\begin{eqnarray}
M_e&=& \frac{24\pi\beta {r_{+}}\left(\beta\left(\ell^2-{r_{+}}^2\right)+4\pi \ell^2{r_{+}}\right)}{[\beta\left(\ell^2-3{r_{+}}^2\right)+4\pi \ell^2{r_{+}}]^2}=\frac{M}{\Xi^2}\,.
\end{eqnarray}
With some algebra it can be proven that these relations confirm the first law 
\begin{eqnarray}\label{firstlaw}
TdS=dm+Jd\Omega+(m+\Omega J)\frac{d\ell}\ell\,,
\end{eqnarray}
In all these expressions we display the dependence on the horizon radius, because otherwise the equations would be very complicated and physical quantities difficult to read off. 

Some interesting consequence of these, admittedly still complicated relations, emerge in large mass limit. For, we see that the asymptotic relations hold
\beq
\be\simeq \frac{4\pi\ell^2}{3r_+}, \quad \Om\simeq\frac{\Xi a}{r_+^2}\,.
\eeq
These  readily imply other asymptotic relations, namely
\beq
m\simeq T^3, \quad F\simeq T^3, \quad PV\simeq T^2\,,
\eeq
which can be seen as AdS/CFT duality at work (in $D$ dimensions the free energy scales as $T^D$). Noteworthy is the entropy formula, which corresponds exactly to the Noether charge formula devised by Wald for asymptotically flat metrics, except for the presence of the absolute value (taken at face value, Wald formula would give a negative entropy in our case). 

We may interpret $m$ as the mass seen by the co-rotating observers at infinity, so that the new mass ${\cal M}=m+\Om J$ would be the mass as seen by a static AdS asymptotic observer. In terms of $\ca M$ the first law takes the conventional form
\beq
TdS=d\ca M-\Om dJ+\ca M\frac{d\ell}\ell\,,
\eeq
or, equivalently,
\beq
\ell TdS=d(\ca M\ell)-\ell\Om dJ\,.
\eeq
That is, on the orbit space of the dilatation group
\[
(\be,\ca M,\Om,J)\to(\la\be,\la^{-1}\ca M,\la^{-1}\Om,J)\,,
\]
where we identify the thermodynamical variables related by a scale transformation, the first law takes the usual form as for rotating black holes in Einstein gravity. This means that moving along an orbit will not change the thermodynamical state of the black hole, in particular, black holes with different masses or angular velocity  can have the same entropy and angular momentum.

The entropy and the angular momentum seem to be the only invariants that characterize the black hole, while the thermodynamical state space is the space of orbits, as discussed extensively in \cite{tn1}.\\
We mention here that asymptotically flat rotating black holes in $R^2$ gravity have vanishing partition function, because $R=0$ on shell. Therefore they have zero energy and entropy, though the temperature may well be different from zero. A possible interpretation of this peculiarity was offered in  \cite{tn1}. 

\section{Conclusion}

\noindent We have generalized to a particular subclass of rotating black holes the results that were obtained in \cite{tn1} on the thermodynamical properties of static and spherically symmetric black holes in $R^{2}$ gravity. We may consider this as a preliminary investigation on the extension to quadratic gravity of the intricate thermodynamical phase structure that AdS black holes exhibit in Einstein gravity. We may note in this respect that the strength of the dark fluid in this theory is not fixed, therefore its actual value can only be fixed by the breaking of scale invariance induced by quantum corrections. 

We have also found equivalent thermodynamical states corresponding to orbits of the dilation group, completely characterized by the entropy and the angular momentum of the black hole (and the charges coupled to gauge fields if these were included).  Apart from this, everything else seems to be similar to what happens in general relativity. The free energy scales with temperature as expected from the AdS/CfT duality, even with the absence of the Einstein term in the action, providing an additional piece of evidence of its general validity. 

\begin{acknowledgments}

\noindent We are grateful to S.\ Zerbini for discussions.

\end{acknowledgments}

{}

\end{document}